\documentstyle[twoside,fleqn,espcrc2,psfig]{article}

\def\arcmin {$^\prime$} 
\def\arcsec{$^{\prime\prime}$} 
\def\msun {\hbox{M$_{\odot}$}}

\title{Three Supernova remnants observed by BeppoSAX}

\author{Jacco Vink\address{SRON-Utrecht, Sorbonnelaan 2 , 3584 CA, Utrecht, The Netherlands}
	}
       
\begin{document}

\begin{abstract}
We present the results of three observations of shell-type supernova remnants
observed by BeppoSAX.
Two of the remnants (N132D and Cas~A) are oxygen rich supernova
remnants. They were observed during the PV phase. SN1006 was
observed during A01. For SN1006 we present preliminary results on the
abundance measurements based on the emission from the center of the remnant.
\end{abstract}

\maketitle

\section{Introduction}

Supernova remnants (SNRs) are of astrophysical interest for a number of 
reasons. Especially young remnants can reveal important information about the 
last stages of stellar evolution; both about the nucleosynthesis as about the
the impact that stellar winds have on the circumstellar medium. In case
of remnants of type Ia supernovae they might help to clarify their origin:
are they the result of the carbon deflagration or detonation of a white dwarf?
From a more physical point of view SNRs are important cosmic laboratories;
displaying phenomena as collisionless shocks, shock acceleration of particles
(the formation of cosmic rays) and the physics associated with tenuous hot 
plasmas. 

For our understanding of SNRs X-ray emission plays an important role;
the bulk of the shock heated plasma can only be seen in X-rays. In that respect
the narrow field instruments of BeppoSAX \cite{Boella97}
(i.e. LECS, MECS, HP, PDS) can play an important
role. As we shall show, especially the broad
energy range and the good sensitivity around 6~keV have already resulted in 
important findings.

In this paper we will give a review of the analysis of BeppoSAX data of
two oxygen rich supernova remnants (N132D and Cas~A) and some preliminary 
results for SN1006.
	
\begin{figure*}
\hbox{
        \psfig{file=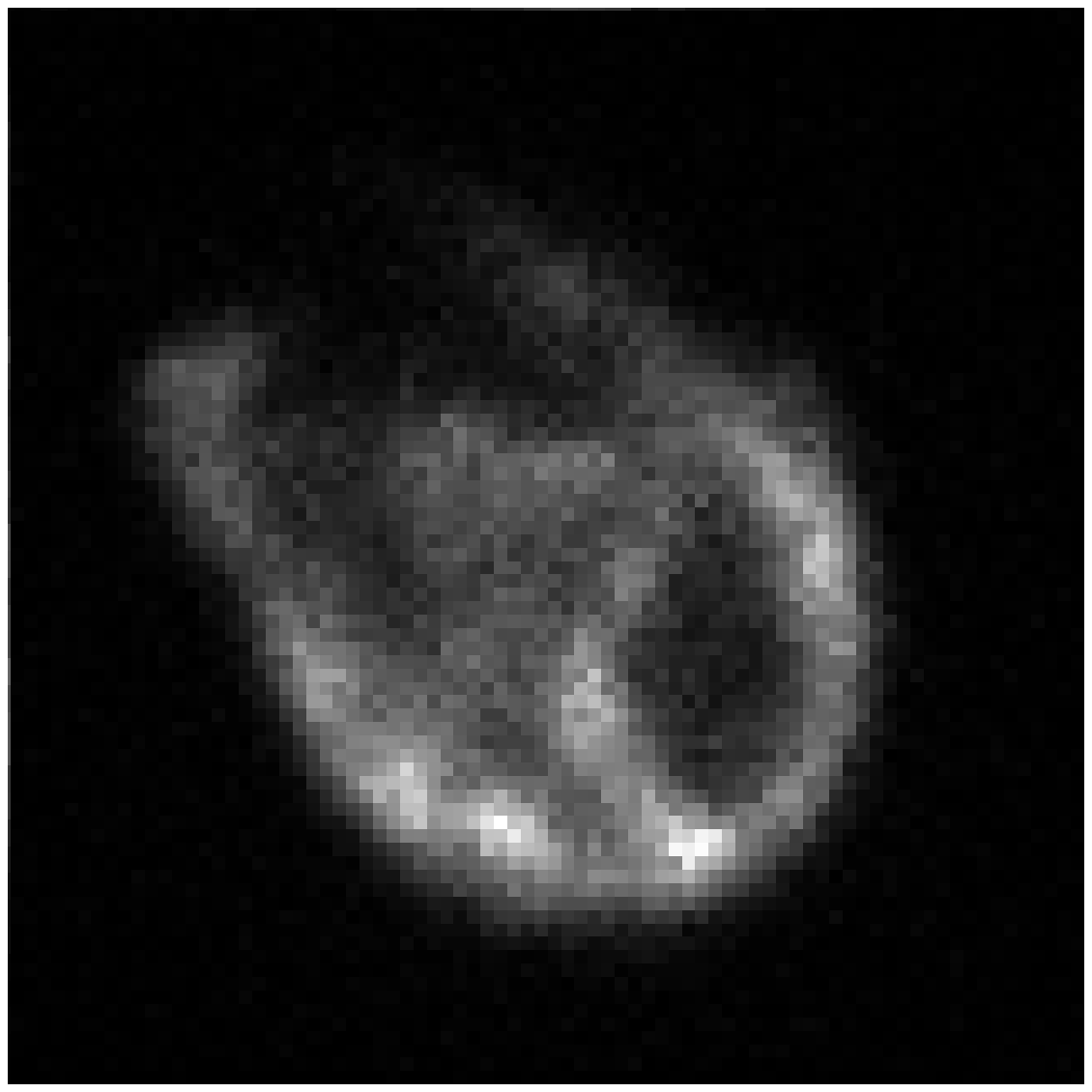,width=7.2cm}
        \psfig{file=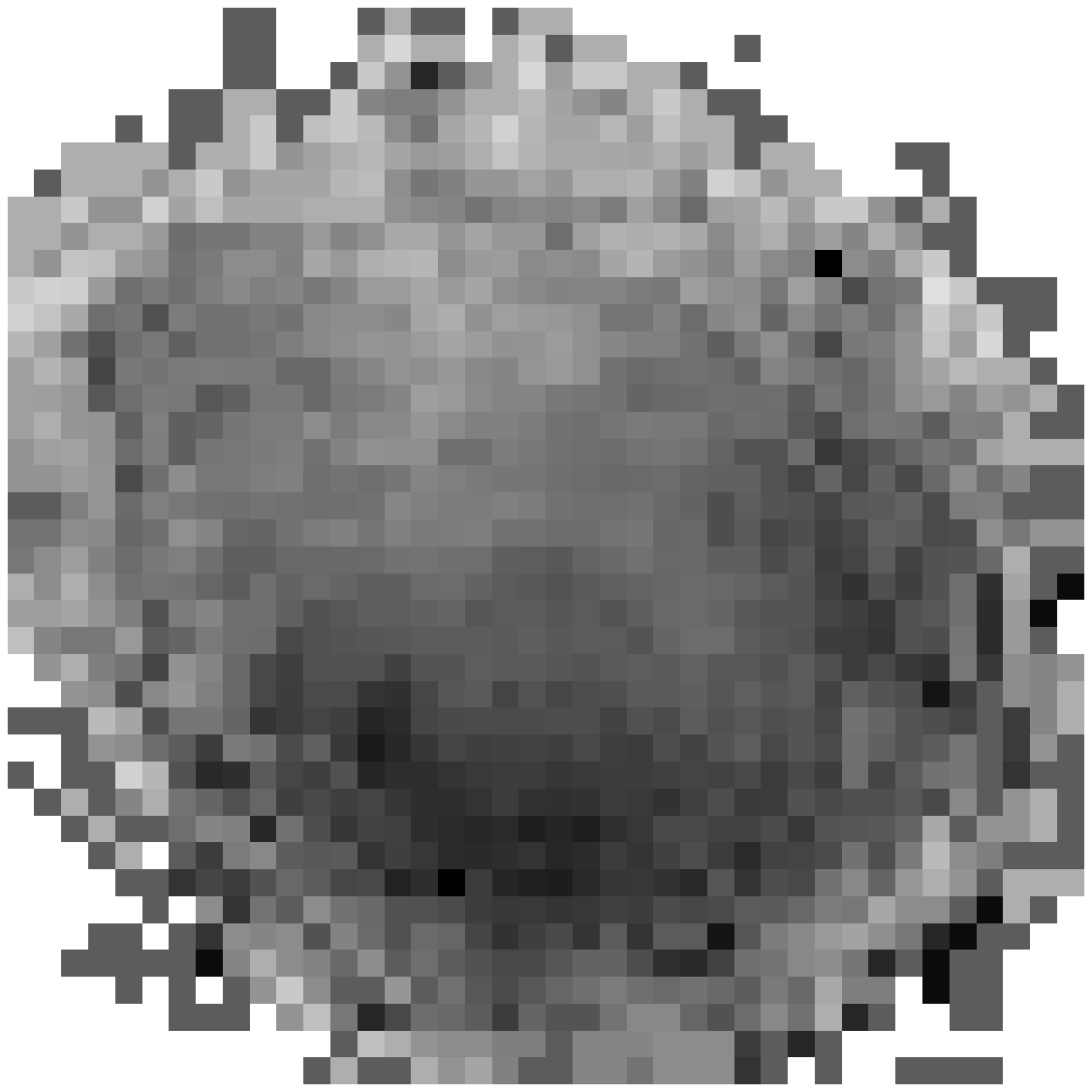,width=7.2cm}
}     
\caption{A ROSAT HRI image (left) and a PSPC ratio image (right) of N132D. 
The ratio image is based on the energy ranges
0.4--0.85~keV and 0.86--2.1~keV, with the ratio varying from 0.5 (soft) 
in the North to 1.5 (hard) in the Southeast. The size of the images is 
2.7\arcmin. The resolution of the PSPC (20\arcsec) is 
worse than the HRI (4\arcsec).\label{N132Dfig}}
\end{figure*}

\section{N132D in the LMC}
Oxygen rich remnants constitute a small but important subclass of SNRs. 
The amount of
oxygen present (often detected first with optical spectroscopy) makes it
very likely that these SNRs are the result of the core collapse of a massive
early type star (an O star or a Wolf-Rayet star). Important examples of this 
subclass are Cas~A, Puppis and N132D. The latter is the brightest remnant in 
the Large Magellanic Cloud (LMC). In fact it is intrinsically brighter than 
Cas~A.
N132D has a size of 1.8\arcmin, so it can not be resolved with
the LECS or MECS. From kinematic and spectroscopic studies of the remnant we
know that its age is about $\sim$ 2500 yrs \cite{Morse}.

\begin{table}
	\caption{
The elemental number abundances with respect to solar abundances for N132D.
\label{N132Dab} 
	}
	\begin{tabular}{lll}
		\hline\noalign{\smallskip}
Element & N132D & LMC \cite{Russell} \\
\noalign{\smallskip}\hline\noalign{\smallskip}
O  & 1.2  & 0.32 \\
Ne & 1.1  & 0.42 \\
Mg & 0.85 & 0.74 \\
Si & 0.62 & 1.7  \\
S  & 0.75 & 0.27 \\
Ar & 0.73 & 0.49 \\
Fe & 0.52 & 0.50 \\
		\noalign{\smallskip}\hline
	\end{tabular}
\end{table}

\subsection{Abundances}
Although optical studies indicate that N132D is an oxygen rich remnant,
a detailed analysis of data from various instruments of the Einstein mission
\cite{Hwang93} indicated underabundances for all major elements with respect to
LMC abundances.  However, an analysis of PV phase data of 
BeppoSAX of N132D \cite{Fav97a} resulted in abundance estimates which were in 
accordance with respect to LMC abundances and even indicated overabundance of 
O, Ne and Mg as can be expected for an oxygen rich remnant 
(see Table~\ref{N132Dab}). 

Another result was that the LECS and MECS indicated the presence of
Fe K emission (around 6.6~keV). As a consequence there must be plasma
present with a substantially hotter electron temperature than the 0.7~keV 
found for N132D. Indeed a spectral fit to the BeppoSAX data using the SPEX
spectral code \cite{Kaastra} gave $kT_{\rm e} = 0.8$~keV and 2.7~keV. 
The emission measure for the hot plasma is almost a factor ten lower than the
cooler plasma, but the hot plasma gives nevertheless a measurable effect.
A significant difference with the Einstein result was that the LECS
indicated an absorption towards N132D of $N_{\rm H} = 3\ 10^{21}$~cm$^{-2}$,
whereas in \cite{Hwang93} a value of  $N_{\rm H} = 6\ 10^{20}$~cm$^{-2}$ 
was listed. It may well be this difference which led to the lower abundances
found for the Einstein data.

\subsection{A hardness map}
Now that we had found that part of the emission comes from a hot plasma, 
we were  naturally interested where the hot plasma might be located. 
As mentioned before, the MECS does not allow to resolve N132D. 
We therefore used archival
ROSAT PSPC data to see if there is any spectral variation over the remnant.
Note however that the PSPC band is still rather soft (up to 2.1~keV), so that
the spectral variations may not be the same as the one that gives rise to 
the hot plasma found in the BeppoSAX spectrum.
Nevertheless, Fig.~\ref{N132Dfig} clearly
shows that there is spectral variation over the remnant, which may be
caused by the interaction of the blast wave with an inhomogeneous 
interstellar medium.

\section{The historical remnant of SN1006}
\begin{figure*}
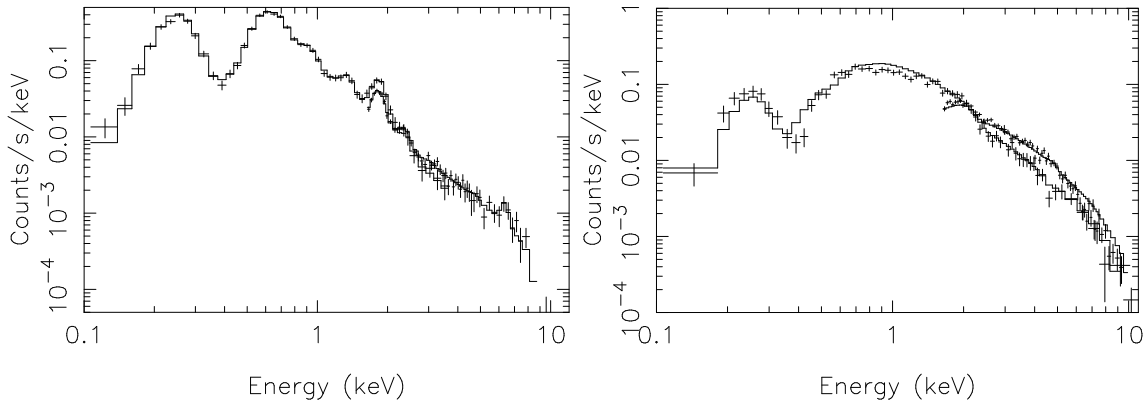

	\centerline{
		\psfig{file=sn1006_Int.ps,width=7.5cm,angle=-90}
		\psfig{file=sn1006_NErim.ps,width=7.5cm,angle=-90}
	}
	\caption{
a) The total LECS/MECS spectrum extracted from the center region of 
SN1006 (the region dominated by thermal emission) 
b) The spectrum of the NE rim.\label{sn1006}
	}
\end{figure*}
\subsection{Two different kinds of spectra}
For a long time the apparent lack of  line features in the X-ray spectrum
of SN1006 was quite puzzling. There is always the possibility that the 
spectrum
was dominated by synchrotron radiation, but SN1006 does not contain a pulsar
like the Crab SNR. The discovery by ASCA \cite{Koyama95} and 
ROSAT \cite{Will95} that the spectrum from the center was thermal 
(i.e. dominated by line emission and bremsstrahlung), but that the bright rims
had a non-thermal spectrum, was a step forward. It was proposed that the
emission from the rims was synchrotron radiation, caused by 
shock-accelerated electrons with energies in excess of TeVs. However,
one should be careful with immediately adapting this point of view,
the arguments are valid, but not conclusive. 
On the other hand, if one assumes, as was done in the past, that some
kind of process suppresses the line emission, one might rightly ask why such
a model does not apply to the emission from the center. In \cite{Will95} it
was reported that a deprojection of the ROSAT PSPC image indicates that
the bright rim are not just the result of limb-brightening, but instead that
they are concentrated in caps. Such a geometry is not easily explained by
simple models for synchrotron emission. See also the discussion below about
the hard X-ray emission of Cas~A.

Another problem associated with SN1006 is that, although it is thought to be
the remnant of a type Ia supernova, there is no sign of the presence of
0.5\msun\ of iron as expected for a type Ia supernova\cite{Hamilton}.

\subsection{Abundances measurements of the central region}

\begin{table}
	\caption{
Elemental mass abundance with respect to oxygen. 
A comparison between the solar values, 
a white dwarf deflagration model \cite{Nomoto} 
and our best fit model for the center of SN1006.\label{sn1006tab}}
	\begin{tabular}{llll}
		\hline\noalign{\smallskip}
Element & Solar & Nomoto & SN1006 \\
\noalign{\smallskip}\hline\noalign{\smallskip}
C & 0.32   & 0.38  & 0.8  \\
N & 0.12   & -     & 0.04 \\
O & 1      & 1     & 1    \\
Ne & 0.18  & 0.03  & 0.12 \\
Mg & 0.067 & 0.06  & 0.14 \\
Si & 0.073 & 1.067 & 1.27 \\
S  & 0.038 & 0.60  & 0.85 \\
Fe & 0.19  & 5.2   & 1.3  \\
		\noalign{\smallskip}\hline
	\end{tabular}
\end{table}

\begin{figure}
	\psfig{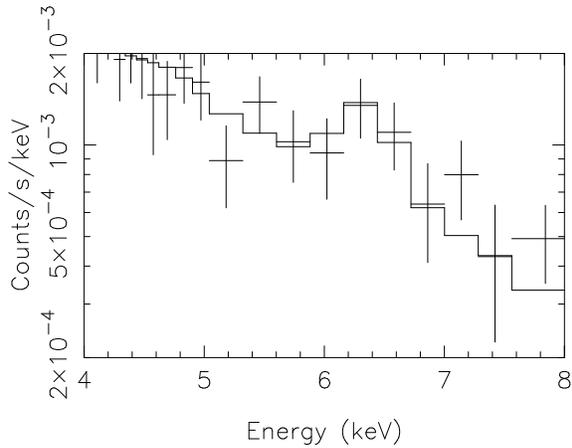}
	\caption{
Detail of Fig.~\ref{sn1006}a showing evidence for Fe K shell emission.
\label{sn1006fe}}
\end{figure}

SN1006 was observed by BeppoSAX in April 1997. The remnant was covered by three
pointings, one at the relatively bright Southeast and two pointings at the
rims. As Fig.~\ref{sn1006} shows the thermal spectrum of the center can
be clearly distinguished from the power law spectrum of the rims. 
Preliminary results from an analysis of the spectrum from the center
are presented in table~\ref{sn1006tab}. The analysis was made with the
spectral fitting program SPEX \cite{Kaastra} and is based on a two component
non-equilibrium ionization model. The ionization parameter found, 
$\log(n_e t) = 9.6$, was in accordance with the age and density 
($\sim 0.1$~cm$^{-3}$) of the remnant.
In the table we compare our spectral fit with the predicted abundances
for a carbon deflagration model and we see that the amount of iron may not be
as high as predicted, but is certainly not far off. Note that our future
analysis should include a proper handling of the spatial behaviour of the 
LECS and the MECS, such as the contamination of the spectrum of the center
by scattered emission from the rims.

An interesting result was that the spectrum of the center shows evidence for
Iron K-shell emission at a $2\sigma$ level with a measured equivalent width of 
($0.6 \pm 0.3$)~keV (Fig.~\ref{sn1006fe}).

\section{Cassiopeia A}
\begin{figure}
	\psfig{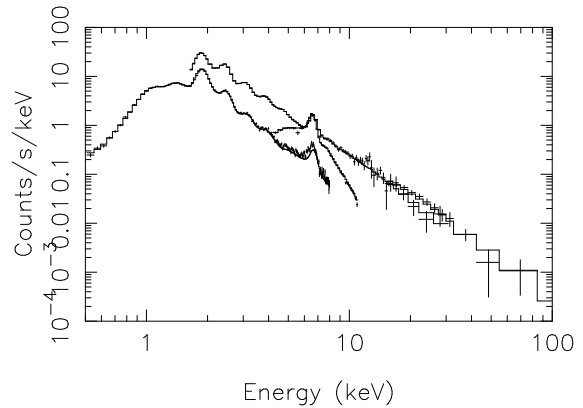}
\caption{The broad-band BeppoSAX spectrum of Cas~A.\label{CasA}}
\end{figure}

\subsection{An oxygen rich remnant with little Ne and Mg}
Cassiopeia A (Cas~A) is as far as we know the youngest (about 310 yrs) remnant
in our galaxy. Given its distance of 3.4~kpc it should have been an 
historical remnant, but no record exist of a bright supernova in the 
second half of the 17th century, although it may have been observed as a 
6th magnitude star by Flamsteed in 1680 \cite{Ashworth}.
Cas~A is the best studied example of an oxygen rich remnant and
it has by now become clear that the progenitor was probably a Wolf-Rayet star 
(WN7 \cite{Fesen87}) with very few hydrogen left in its envelope. 
As a consequence, part of the hot plasma may be hydrogen poor which leads to 
a lower mass estimate for the ejecta, namely around 4\msun\ \cite{VKB96}. 
The reason is that the most abundant element in the ejecta, oxygen, 
is a more efficient bremsstrahlung emitter than Hydrogen. The mass estimate is
based on the assumption that the ejecta, shocked by the reverse shock, are
associated with the low temperature component ($\sim$ 0.7~keV), 
whereas the shocked circumstellar medium has a temperature of $\sim$4~keV.

Although well studied there are some unanswered questions with respect to 
Cas~A.
One is about the nature of the stellar remnant that the 
core-collapse must have left behind. 
Despite several attempts at various wavelength
no stellar remnant has been discovered. So can we conclude that
Cas~A contains a black hole? Possibly, but this seems at odds with the 
recent discovery of emission from radio-active $^{44}$Ti by the CGRO/Comptel 
experiment \cite{Iyudin,Dupraz}, as we will explain below. 
A puzzling fact is also the rather small amount of neon and 
magnesium \cite{VKB96}, which is not expected for an oxygen rich remnant.
The lack of Ne and Mg was confirmed by the BeppoSAX data \cite{Fav97b}.

\subsection{The hard tail}
Cas~A was observed by the BeppoSAX narrow field instruments during the PV 
phase. The most remarkable finding was the existence of
a hard tail to the spectrum up to 60~keV \cite{Fav97b}. 
This was not quite unexpected,
since it was already detected by HEAO-2 \cite{Pravdo} and confirmed by
the CGRO/OSSE experiment \cite{The96}. The hard tail was also detected
by RossiXTE \cite{Allen,Roth}. These studies indicate 
that the hard-tail is best fitted with a power law spectrum with a 
photon index close to 3. 

So there is agreement that such a hard tail exist; 
however, there is no agreement yet on the nature of the emission.
There are several possible emission mechanism, the most likely ones
being (non-thermal) bremsstrahlung and synchrotron emission. 
The synchrotron emission was already discussed in connection to SN1006.
In the case of Cas~A, with its magnetic field in the order of 1mG, 
it would indicate the presence of shock accelerated electrons with energies 
up to 40~TeV \cite{Allen}. On the other hand, a bremsstrahlung model needs only
energies of several tens to hundreds of keV. Such electron energies may be the
signature of electrons that have just been accelerated by shocks or they may
arise from electrons interacting with plasma waves. Note that we can be sure 
that Cas~A should emit bremsstrahlung in the PDS band, simply because we
know from the synchrotron emission observed at radiowavelength that there 
exists a population of electrons with energies in the MeV to GeV range. Since
electrons are continuously accelerated from thermal energies there must 
also exist a population of electrons emitting bremsstrahlung at intermediate
energies. The reason that some discard this explanation for the hard 
tail is that a simple calculation of this so-called electron injection spectrum
predicts a photon-index of 2.26 \cite{Asvarov}. However, 
things are not so easy; the injection of electrons in a shock are poorly 
understood and the actual energy spectrum 
may very well be consistent with the observed photon index \cite{Bykov}.
Finally, a preliminary analysis of the MECS data indicates that the 
iron K-shell emission has a different distribution than the high energy 
continuum \cite{Maccarone}. This may favour the synchrotron model, but it may
also indicate that the iron emission does not completely trace the hot plasma, 
simply because the iron is not present everywhere with equal abundances.

\subsection{Abundances}
The nature of the hard tail is important for the modeling of the overall
X-ray spectrum of Cas~A. If the hard tail is mainly the result of 
bremsstrahlung, we can trust the abundances derived from the line
intensities below 10~keV. However, if the mechanism is synchrotron we
are dealing with a very different continuum, also at lower energies. The line
to bremsstrahlung continuum ratio is then much higher than in the former case,
resulting in higher abundance estimates and lower inferred densities and 
plasma mass. 
A comparison between abundances listed in \cite{Fav97b} reveals
that treating the continuum to be partly synchrotron radiation gives abundances
a factor 3 higher and a mass estimate which is 40\% lower
($\sim$2\msun\ instead of $\sim$4\msun\ of shocked ejecta).

\subsection{Titanium 44}
During the core collapse of a star heavy elements are formed, some of them
are radio-active such as $^{56}$Ni (6.1 days half-life).
Another important radio-active isotope is  $^{44}$Ti, which has a half-life
time of about 50 yrs and decays in  $^{44}$Sc (6hr half-life), 
which decays into $^{44}$Ca.  
The decay of $^{44}$Sc was detected by CGRO/Comptel at 1.157~MeV
\cite{Iyudin,Dupraz}. 
During its decay to  $^{44}$Sc   $^{44}$Ti
emits two photons at 67.9~keV and 78.4~keV. 
The flux in theses lines should
correspond to the flux at 1.157~MeV of $(3.4\pm0.9) 10^{-5}$cm$^{-2}$s$^{-1}$ 
reported for Comptel \cite{Dupraz}. 
Attempts to measure these lines with
RossiXTE \cite{Roth} and CGRO/OSSE \cite{The96} are not yet successful, 
although the results are not inconsistent with the Comptel results.

The importance of  $^{44}$Ti is that it is directly linked with processes
during the core-collapse. A substantial amount of the $^{44}$Ti 
is therefore inconsistent with the presence of a massive black-hole in Cas~A.
For the same reason the
lack of iron 56 (a product of nickel 56) is an indicator for the presence of
a black hole, so this seems at odds with the presence $^{44}$Ti. However,
a solid detection of $^{44}$Ti may provide an opportunity to finetune
the current models and give us new insights in the details
of the explosion or even allow to obtain a mass estimate for the
black-hole that may be present in Cas~A \cite{Nagataki}.

We also made an attempt to measure the line flux with the PDS, 
but up to now we can only come with an upper limit of  
$5\ 10^{-5}$cm$^{-2}$s$^{-1}$ (99\% confidence) for 35ksec of data. 
In the near future when more data will be available
we hope to come with a more substantial result.

\section{Concluding remarks}
We have shown the potential of the BeppoSAX narrow field instruments. The
broad energy range allows for the detection of hard energy tails 
(such as in Cas~A) and a better determination of the absorption at low energies
(as in SN1006). Both are essential for a better understanding of the X-ray 
spectra of SNRs. Furthermore the MECS sensitivity around 6.5~keV resulted
in the detection of Fe K complexes not yet seen with other missions 
(SN1006 and N132D). The Fe K complex is an important diagnostic 
tool for determining hot plasma parameters such as temperature and ionization 
stage.
Note that the sensitivity is as much due to a reasonable
effective area as to a low background; the latter being the result of a 
telescope mirror of good quality. 

\section*{Acknowledgments}
I would like to thank Jelle Kaastra, Rolf Mewe and Johan Bleeker 
for their support and 
enthusiasm for this research. Part of the results presented here have already
been published, I would therefore like to thank all authors of the articles 
\cite{Fav97a,Fav97b} for their contributions. Finally, I would like to thank 
Glenn Allen for discussions on the hard X-ray emission of Cas~A.

\end{document}